\documentclass[twocolumn,preprintnumbers,amsmath,amssymb,superscriptaddress]{revtex4}
\usepackage{epsfig,bm}
\usepackage{ulem}
\usepackage{amsmath,color}

\begin{document}
\title{Emergence of nuclear clustering in 
electric-dipole excitations of $^6$Li}
\author{S. Satsuka}
\affiliation{Department of Physics,
  Hokkaido University, Sapporo 060-0810, Japan}
\author{W. Horiuchi}
\affiliation{Department of Physics,
  Hokkaido University, Sapporo 060-0810, Japan}

\begin{abstract}
  Nuclear clustering plays an important role, especially in the dynamics
  of light nuclei. The importance of the emergence of the nuclear clustering 
 was discussed in the recent measurement of the photoabsorption cross sections
 as it offered the possibility of
the coexistence of various excitation modes
which are closely related to the nuclear clustering.
To understand the excitation mechanism, we study the electric-dipole
  ($E1$) responses
  of $^6$Li with a fully microscopic six-body calculation.
  The ground-state wave function is accurately described with a superposition
  of correlated Gaussian (CG) functions with the aid of
  the stochastic variational method. The final-state wave functions
  are also expressed by a number of the CG functions including
  important configurations to describe the six-body continuum states
  excited by the $E1$ field.
  We found that the out-of-phase transitions occur due to the oscillations
  of the valence nucleons
  against the $^4$He cluster at the low energies around 10 MeV
  indicating ``soft'' giant-dipole-resonance(GDR)-type excitations,
  which are very unique in the $^6$Li system
  but could be found in other nuclear systems.
  At the high energies beyond $\sim 30$ MeV typical GDR-type transitions occur.
  The $^3$He-$^3$H clustering plays an important role to the GDR phenomena
  in the intermediate energy regions around 20 MeV.
\end{abstract}
\maketitle

\section{Introduction}

Nuclear cluster structure often appears in the spectrum of light nuclei.
Especially, the $\alpha$($^4$He) cluster plays
a vital role to explain the low-lying spectra
of $N=Z$ nuclei~\cite{Ikeda68,Fujiwara80}.
Much attention has been paid to the understanding
for the role of the nuclear clustering
in the electromagnetic transitions of light nuclei 
as their importance in the nucleosynthesis
represented by the triple $\alpha$ processes
related to the famous Hoyle state in $^{12}$C~\cite{Hoyle}.

The electric-dipole ($E1$) transition strengths
contain numerous information 
on the ground- and final-state wave functions
and have often been used as a probe 
for the the nuclear structure and dynamic properties. 
The giant dipole resonance (GDR)
can be observed in any nuclear systems,
which has been recognized as the classical picture
of the out-of-phase oscillation
between protons and neutrons induced by the $E1$ external field
~\cite{Goldhaber48, Steinwedel50}.
Since its resistance force stems from the nuclear saturation properties,
the peak position as well as its distributions is closely related
to the bulk properties of the nuclear matter, 
especially, to the nuclear symmetry energy~\cite{Horowitz01, Tamii11}

Recently, due to the new advancement of the experimental techniques,
the exploration of new $E1$ excitation modes has attracted interest
to the nuclear physics community. In neutron-rich nuclei
in which the neutron wave function is extended than that of protons, 
it was pointed out that in the low excitation energies
the possibility of emerging the soft dipole excitation mode
as oscillations of a core against
excess neutrons~\cite{Hansen87, JHP87, Ikeda88, Suzuki90}.
Recent microscopic calculations for halo nuclei showed that
that the low-lying $E1$ strengths have the typical soft-dipole type
excitation character in $^{6}$He ~\cite{Mikami14} and $^{22}$C~\cite{Inakura14}.
Exotic excitations such as troidal and compressive modes
were also discussed in $^{10}$Be as possible new excitations
for light unstable nuclei~\cite{Enyo17}.

Recently, the photoabsorption reaction cross sections
of $^6$Li were measured~\cite{Yamagata17}
in the energy range up to $\sim 60$ MeV,
where the $E1$ transitions are dominant.
A two-peak structure in the photoabsorption cross sections
was found and the peak at the lower and higher energies 
were respectively conjectured
as the GDR of $^6$Li and the GDR of the $\alpha$ cluster
in $^6$Li based on the idea given in
the early study of the $^{6}$Li($\gamma,n$)
reaction~\cite{Costa63}
for the interpretation of the higher peak.
If this interpretation is true,
these kinds of subnuclear excitation modes
may appear in various nuclear systems
where the $\alpha$-cluster structure is well developed. 

In this paper, we study the $E1$ transitions of $^{6}$Li.
The $^{6}$Li nucleus has often been described with an $\alpha+p+n$
three-body model (See, for example~\cite{Horiuchi07,Watanabe15} and references
therein). However, to understand the $E1$ excitation mechanism of $^6$Li,
a fully microscopic six-body calculation is needed
that can describe the formation and distortion of nuclear clusters
in a wide range of the excitation energies up to $\sim$60 MeV.
We calculate the $E1$ transition
strengths and their transition densities
and discuss how $^6$Li is excited by the $E1$ field as a function
of the excitation energy. We discuss the roles of light clusters 
in the $E1$ excitation spectrum extending the discussions
given in Ref.~\cite{Mikami14}.
In that paper, the proton-proton distance in
the wave function of $^6$He was
introduced as a measure of the $\alpha$ clustering.
However, it is not useful for the case of $^6$Li because
the wave function is totally antisymmetric with respect to
the exchange of the nucleons and thus
we cannot distinguish protons either in the $\alpha$ cluster
or the valence part of $^6$Li.
Therefore, we calculate the spectroscopic factors of various cluster
configurations as direct quantities of the clustering degrees-of-freedom.

The paper is organized as follows. In the next section,
we define basic inputs used 
in the many-body variational calculation.
Section~\ref{wave.sec} explains details of the procedures to obtain
the ground- and final-state wave functions 
for the six-nucleon system.
In Sec.~\ref{results.sec}, we calculate the $E1$ transition
strength distributions
and discuss how the transition occurs as a function
of the excitation energy by analyzing their transition densities
focusing on the roles of the nuclear clustering.
The compressive isoscalar dipole strengths are also evaluated
as they reflect the other profiles of the transition densities.
Conclusion is made in Sec.~\ref{conclusion.sec}.

\section{Microscopic few-nucleon model}
\label{model.sec}

Here we briefly describe the microscopic few-nucleon
model employed in this paper.
The Hamiltonian for an $N$-nucleon system consists of
the kinetic energy and two-body potential terms $V_{ij}$ as
\begin{align}
  H=\sum_{i=1}^{N}T_i-T_{\rm cm}+\sum_{i>j}V_{ij},
\end{align}
where $T_i$ is the kinetic energy of the $i$th nucleon.
The center-of-mass (cm) kinetic energy $T_{\rm cm}$ is
properly subtracted, and hence no spurious $E1$ excitation
appears in the calculation. 
As a nucleon-nucleon potential, we employ an effective central potential, 
the Minnesota (MN) potential~\cite{MN} which
offers a fair description of the binding energies
and radii of $s$-shell nuclei, $^2$H ($d$), $^3$H ($t$), $^3$He ($h$), 
and $^4$He ($\alpha$)~\cite{Varga95, Suzuki08} without a three-body force.
The MN potential includes the one parameter $u$ that controls the strength
of the odd-parity waves. Later we will discuss how the results depend on
the $u$ parameter. The Coulomb interaction is also included.

The $N$-nucleon wave function with spin $J$ and its projection $M_J$ are 
expanded in terms of the fully antisymmetrized
basis function $\Phi_{JM_J,i}^{(N)}$
\begin{align}
  \Psi_{JM_J}^{(N)}=\sum_{i=1}^KC_i^{(N)}\Phi_{JM_J,i}^{(N)}
\label{wfdef.eq}
\end{align}
For the basis function, we employ the global vector representation
of the correlated Gaussian basis function~\cite{Varga95, SVM} as
\begin{align}
  \Phi_{JM_J,i}^{(N)}=
\mathcal{A}\left\{
  \exp\left(-\frac{1}{2}\tilde{\bm{x}}A_i\bm{x}\right)
  \left[\mathcal{Y}_{L_i}(\tilde{v}_i\bm{x})\chi_{S_i}^{(N)}\right]_{JM_J}\eta_{M_T}^{(N)}\right\},
\label{CG.eq}
\end{align}
where $\mathcal{A}$ is the antisymmetrizer.
As the coordinate set $\bm{x}=(\bm{x}_1,\dots,\bm{x}_{N-1})^t$ 
excluding the cm coordinate of the $N$-nucleon system, $\bm{x}_N$,
we conveniently take it as the Jacobi coordinate set:
$\bm{x}_k=\bm{r}_{k+1}-\bm{x}_{\rm cm}^{(k)}$
with the cm coordinate of $k$-nucleon
subsystem, $\bm{x}_{\rm cm}^{(k)}=\sum_{j=1}^k r_j/k$.
A tilde denotes the transpose of a matrix.
The matrix $A$ is an $(N-1)$-dimensional positive-definite symmetric matrix.
The correlations among the particles
are explicitly described through the off-diagonal matrix elements
of $A$ noting that 
a quadratic form $\tilde{\bm{x}}A\bm{x}=\sum_{jk}A_{jk}\bm{x}_j\cdot\bm{x}_k$
on the exponent. The rotational motion of the system
is described with the so-called global vector 
$\tilde{v}\bm{x}=\sum_{j=1}^{N-1}v_j\bm{x}_j$~\cite{Suzuki98,SVM}.
Because the functional form does not change under
any linear transformation of the coordinate,
the form of Eq. (\ref{CG.eq}) is convenient to include
various configurations such as 
single-particle, $\alpha+p+n$ and $h+t$ cluster configurations 
being described in Sec.~\ref{final.sec}.
With this nice property,
the correlated Gaussian method has been applied to many examples
related to the nuclear clustering. The readers are referred to
various applications~\cite{Horiuchi08, Horiuchi14, Ohnishi17}
and review papers~\cite{Mitroy13, Suzuki17}.

The $N$-nucleon spin function with the total spin $S$ and its projection
$M_S$ is given as the successive coupling 
of $N$ single-particle spin functions $\chi_{\frac{1}{2}m_s}$ as 
\begin{align}
\chi_{SM_S}^{(N)}=
\left[\dots\left[\left[\chi_{\frac{1}{2}}(1)\chi_{\frac{1}{2}}(2)\right]_{S_{12}}\chi_{\frac{1}{2}}(3)\right]_{S_{123}}
\dots\right]_{SM_S}.
\end{align}
All possible intermediate spins $S_{12}, S_{123}, \dots, S_{123\cdots N-1}$ 
are taken into account in the calculation.
The isospin function with the total isospin $T$ and its
projection $M_T=\sum_{j=1}^N m_{t_j}$ is represented by 
the particle basis which is the direct product of
$N$ single-particle isospin functions $\eta_{\frac{1}{2}m_t}$ as
\begin{align}
\eta_{M_T}^{(N)}=\eta_{\frac{1}{2} m_{t_1}}(1)\cdots\eta_{\frac{1}{2} m_{t_N}}(N)
\end{align}
with $m_{t_j}=1/2$ for neutron and $-1/2$ for proton.
In the particle basis, the mixture of possible total isospin states
with, e.g., $T=0, 1, 2,$ and 3 for $^6$Li, is naturally taken into account.

After those parameters of the basis functions are set,
we determine the $K$-dimensional coefficient vector $\bm{C}=(C_1^{(N)},\dots,C_K^{(N)})^t$
by solving the generalized eigenvalue problem
\begin{align}
  H\bm{C}=EB\bm{C}
\end{align}
  with $B_{ij}=\left<\Phi_{JM_J,i}^{(N)}|\Phi_{JM_J,j}^{(N)}\right>$ and
  $H=\left<\Phi_{JM_J,i}^{(N)}\right|H\left|\Phi_{JM_J,j}^{(N)}\right>$.
  These matrix elements can be evaluated analytically.
  See~\cite{Varga95, SVM, Suzuki08} for
  the detailed mathematical derivation and expressions.

\section{Calculations of the wave functions}
\label{wave.sec}

In this paper, we mainly discuss the $E1$ transitions.
The reduced $E1$ transition probabilities or $E1$ transition strengths
are defined by
\begin{align}
  &B(E1,E_f)\notag\\
  &=\frac{1}{2J_0+1}\sum_{J_fM_fM_0\mu}
\left|\left<\Psi_{J_fM_f}^{(N)}(E_f)
\right|\mathcal{M}(E1,\mu)\left|\Psi_{J_0M_0}^{(N)}(E_0)\right>\right|^2
\end{align}
with the $E1$ operator 
\begin{align}
\mathcal{M}(E1,\mu)=
e\sqrt\frac{4\pi}{3}\sum_{i\in p}^{N}\mathcal{Y}_{1\mu}(\bm{r}_i-\bm{x}_{N})
\end{align}
with a solid spherical harmonic,
$\mathcal{Y}_{\lambda\mu}(\bm{r})=r^\lambda Y_{\lambda\mu}(\hat{\bm{r}})$,
where the summation $i$ runs only for proton.
In this section, we describe detailed setup of the calculations
for the initial-ground $\Psi_{J_0M_0}(E_0)$ and
final-continuum $\Psi_{J_fM_f}(E_f)$ state wave functions.

\subsection{Ground-state wave functions}
\label{ground.sec}

For the ground-state wave function with the total angular
momentum and parity $J^\pi=1^+$, in this paper,
we consider only the total orbital angular momentum
$L=0$ with the total spin $S=1$ state
because the MN potential does not mix
with the higher angular momentum states.
It does not mean that the particles are not correlated.
Higher partial waves for all relative coordinates
are taken into account through the off-diagonal
matrix elements of the matrix $A$ of Eq.~\ref{CG.eq}
in the optimization procedure
explained below.

As mentioned in the previous section,
we need to optimize a huge number of the variational parameters.
To achieve it efficiently, we employ the stochastic variational method 
(SVM)~\cite{Varga95,SVM}.
First we adopt the competitive selection
from randomly selected candidates
and increase the basis size until a certain number of basis states
is obtained with $u=1.00$. 
Then we switch the selection procedure
for the refinement of the variational parameters in
the already obtained basis functions
until the energy is converged within tens of keV.
The convergence is reached at $K=600$ in Eq.~(\ref{wfdef.eq})
as adopted in Ref.~\cite{Mikami14}.
This number is very small by noting that
the matrix $A$ includes $N(N-1)/2$ parameters
as well as the spin degrees of freedom for each basis function.
For the wave functions with other $u$ parameters,
we start with the optimal basis functions with $u=1.00$ and
refine those basis functions by keeping the total number of
basis unchanged until the energy convergence is reached.

Table~\ref{gnd.tab} lists the ground-state properties of $^6$Li
with different values of the $u$ parameter in the MN potential.
As shown in the binding energy of $^6$Li, $E_0(^6$Li),
the original MN potential ($u=1.00$) offers little too strong
odd-wave interaction to reproduce
the two-nucleon separation energy of $^6$Li, $S_{pn}$,
leading to the small rms point-proton radius, $r_{p}$,
compared to the measurement~\cite{Angeli13}.
It is noted that the $u$ parameter does not affect
the interaction for the even-parity partial waves
but only for the odd-parity ones.
Roughly speaking, the $u$ parameter controls the interaction of
the valence nucleons from the $\alpha$ core on the $p$-shell orbital. 
In fact, as listed in Table~\ref{gnd.tab},
the binding energies of the $\alpha$ particle,
$E_0(\alpha)$, have almost no dependence on $u$.
Therefore, we prepare two more sets by
considering the repulsive odd-wave strength: One set is to
reproduce the experimental $S_{pn}$ value ($u=0.93$),
and the other set reproduces the experimental $r_p$ value ($u=0.87$).
As shown in the table,
the smaller the $S_{pn}$ value is, the larger the nuclear radii becomes.
Little difference between $r_p$ and the rms point-neutron radius, $r_n$, 
is due to the Coulomb interaction, which is naturally described in
this study. The proton-proton distance, $r_{pp}$, is also
calculated and listed in the table for the sumrule evaluation
which will be discussed later.

\begin{table}[hbt]
\begin{center}
  \caption{Ground-state properties of $^6$Li. Energy and radii are
    given in units of MeV and fm, respectively. See text for details.
    The experimental data is taken from Refs.~\cite{Tilley02,Angeli13}.}
  \begin{tabular}{ccccccccc}
      \hline\hline
$u$&$E_0 (^6$Li) &$E_0 (\alpha)$&$S_{pn}$
&$r_m$&$r_p$&$r_n$&$r_{pp}$&$S_{\alpha d}^2$\\
\hline
1.00&$-34.63$&$-$29.94&4.7&2.20&2.20&2.20&3.62&0.856\\
0.93&$-33.63$&$-$29.90&3.7&2.33&2.34&2.33&3.86&0.869\\
0.87&$-32.94$&$-$29.87&3.1&2.45&2.46&2.45&4.07&0.882\\
\hline
Expt.&$-$31.99&$-$28.30&3.70&    &2.452&\\
\hline\hline
  \end{tabular}
  \label{gnd.tab}
\end{center}
\end{table}

\begin{figure}[ht]     
  \begin{center}
    \epsfig{file=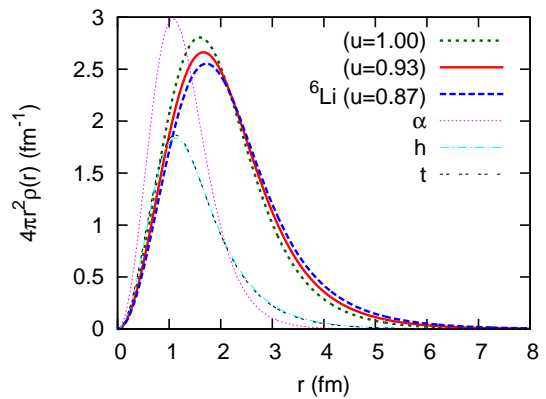, scale=1.2}    
    \caption{Point-matter densities of $^6$Li
with different values of the $u$ parameter in the MN potential.
The densities of $\alpha$, $h$, and $t$
with $u=0.93$ are also plotted for comparison.}
    \label{dens.fig}
\end{center}    
\end{figure}

Figure~\ref{dens.fig} plots the point-matter
densities of $^{6}$Li with different values of the $u$ parameter.
The nuclear surface slightly extends with decreasing the odd-wave strength
or the $u$ parameter.
Peak positions are located in the range 1.6-1.7\,fm
and their magnitude become half at about 2.7-2.8 fm. 
We also plot the densities of $\alpha$,
$h$, and $t$ with $u=0.93$.
The calculated binding energies of $h$ and $t$ are respectively
$-$7.68 and $-$8.38 MeV for $u=1.00$.
They also do not depend on the choice of the $u$ parameter.
Only 0.01 MeV difference is obtained for decreasing
$u$ to $0.93$ and $0.87$.
The peak of the density of $\alpha$ is at about 1\,fm, showing
the sharper distribution as compared to that of $^6$Li.
The peak position of $h$ and $t$ are almost the same as
that of $^4$He but the half-density positions are somewhat
larger than that of $^4$He due to the weaker binding.

The $^6$Li nucleus is known to have developed $\alpha$ cluster structure
and is well described with
an $\alpha+p+n$ three-body model~\cite{Horiuchi07}
having significant amount of the $\alpha+d$
component~\cite{Watanabe15,Kawamura19}.
As a measure of the clustering degrees-of-freedom,
we also show the spectroscopic factor of
the $\alpha+d$ configuration. The probability of finding
the two-cluster ($a$ and $b$) configurations
in the $^6$Li wave function is defined by
\begin{align}
  S_{ab}^2&=  \left|
  \left<\left.\Psi^{(a)}\Psi^{(b)}
  \right|\Psi^{(6)}_{JM_J}(E)\right>\right|^2,
\label{SFab.eq}
\end{align}
where $\Psi^{(a)}$, $\Psi^{(b)}$, and $\Psi^{(6)}_{JM_J}(E)$
are the ground-state wave functions
of nuclei $a$, $b$, and $^6$Li 
with energy $E$, respectively.
The relative wave function on
the coordinate between the cm of the nuclei $a$ and $b$
is integrated out. Details of this evaluation is given in Appendix.
We calculate the $S_{\alpha d}^2$ values ($a=\alpha$, $b=d)$ for
the ground-state wave functions which are listed in Table~\ref{gnd.tab}.
The $S_{\alpha d}^2$ values are found being
large 0.86--0.88 for all the values of the $u$ parameter,
which is consistent with the value obtained with
the variational Monte Carlo calculation, 0.84~\cite{Forest96}.
The $\alpha+d$ cluster structure is somewhat
distorted by the $NN$ interaction and the Pauli principle.
In fact, the $S_{\alpha d}^2$ value slightly increases
by adding the repulsive $\alpha$-nucleon interaction
with $u=0.93$ and $0.87$.
Though $S_{\alpha d}^2$ values are large,
no bound state in the $^6$Li system is obtained
only with the $\alpha+d$ configurations with a relative $s$-wave.
Inclusion of those distorted configurations is essential
to get binding in the six-nucleon system.

\subsection{Final-state wave functions}
\label{final.sec}

\begin{figure}[ht]     
  \begin{center}
\epsfig{file=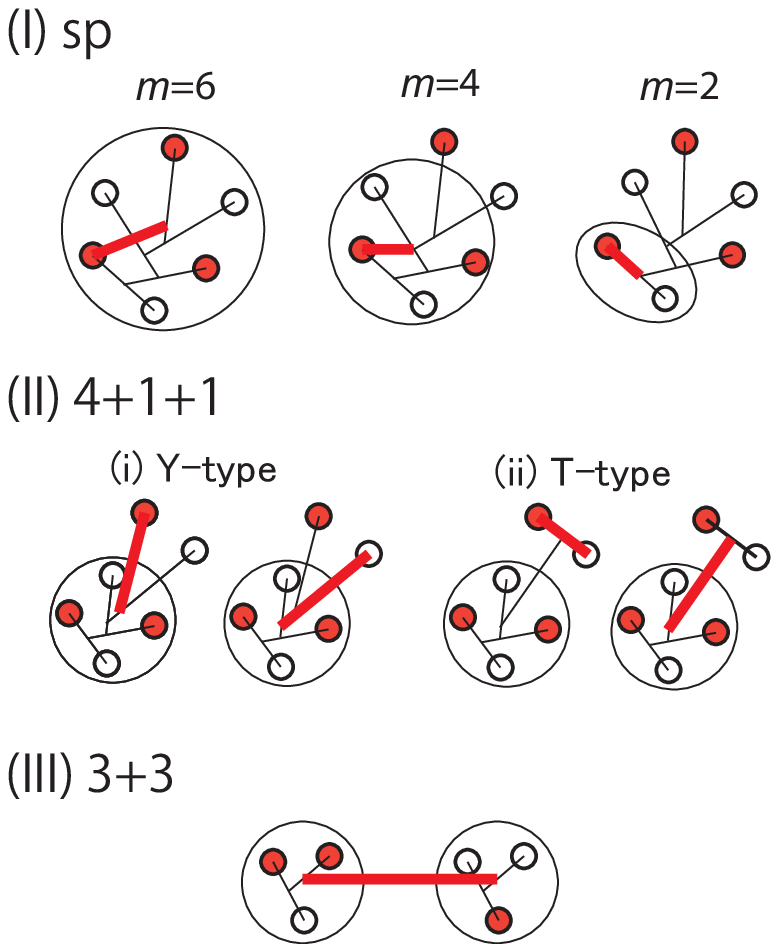, scale=1}    
\caption{Schematic figures of the basis functions for the final-state
  wave functions. Colored and uncolored small circles represent
  protons and neutrons, respectively.
  Thick lines indicate the coordinates excited by the $E1$ operator.
  See text for exact definitions.}
    \label{channels.fig}
\end{center}    
\end{figure}

In this subsection, we explain how 
to construct the final-state wave functions excited by the $E1$ operator.
We follow almost the same prescription as given
in Refs.~\cite{Horiuchi12,Horiuchi13,Mikami14}
but are extended for adopting to the $^6$Li case.
The ground-state wave function with $J^\pi=1^+$
is excited by the $E1$ operator.
Since the operator does not change the spin of the initial state $S=1$,
the orbital angular momentum of the final state should be $L=1$
and thus $J^\pi=0^-, 1^-$, and $2^-$ states need to be considered.
In this paper, we did not include the spin-orbit interaction.
These states are energetically degenerate
that only its multipolarity is different.

We expand the final-state wave function in a large number
of the correlated Gaussian basis functions of Eq. (\ref{CG.eq}).
To incorporate the complicated six-body correlations efficiently,
physically important configurations are selected and categorized into
the three types below:
(I) Single-particle (sp) excitation,
(II) 4+1+1 cluster, 
and (III) 3+3 cluster configurations.
All these configurations are again expressed by
the same functional form as of Eq. (\ref{CG.eq})
with appropriate coordinate transformations.

The configurations of type (I) is based on the idea that
the $E1$ operator excites one coordinate
in the ground-state wave function and these configurations
are further subcategorized into three channels which will
be explained later in this paragraph.
The resulting coherent states are important
to satisfy the $E1$ sumrule~\cite{Horiuchi12}.
The configurations of type (I) are constructed by
using the basis set of the ground-state wave function of $^6$Li by multiplying
additional angular momentum $L=1$ as
\begin{align}
\Phi_{JM_J,i}^{{\rm (sp,}n)}=\mathcal{A}\left[\Phi_{1,i}^{(6)}(123456)\mathcal{Y}_{1}(\bm{r}_1-\bm{x}_{\rm cm}^{(m)})
\right]_{JM_J},
\end{align}
where $\Phi_{1,i}^{(6)}$ is the $i$th basis $(i=1,\dots, 600)$ 
of the ground-state wave function of $^6$Li.
The coordinate $\bm{r}_1$ denotes the single-particle coordinate of a proton.
As a first choice, we take the coordinate of one proton 
measured from the cm of the system ($m=6$).
Considering that $^6$Li has large $S_{\alpha d}^2$ component $\sim$0.9,
we include the additional sp basis sets that the four- and two-nucleon
subsystems are excited by the $E1$ operator (the channels with $m=4$ and 2).
Finally, the total number of the basis of the type (I) is 1800
including the channels with $m=6, 4$, and 2.

The configurations of types (II) and (III) explicitly take care of
the cluster configurations of $\alpha+p+n$ and $h+t$,
which correspond to the two lowest thresholds,
3.7 and 15.8 MeV~\cite{Tilley02}, respectively.
They are expected to be important for describing
the low-lying ($\lesssim$16 MeV) and intermediate energies ($\gtrsim$16 MeV).

The configurations of type (II) are defined in the following
\begin{align}
&\Phi_{JM_J,ijk}^{(411,l)}=\mathcal{A}\left\{\Phi_{0,i}^{(4)}(1234)
\exp\left(-\frac{1}{2}\tilde{\bm{y}}B^{(jk)}\bm{y}\right)\right.\notag\\
&\times
\left.\left[\mathcal{Y}_1(\bm{y}_l^{({\rm X})})
\left[\chi_{\frac{1}{2}}(5)\chi_{\frac{1}{2}}(6)\right]_{S_{56}}\right]_{JM_J}
\eta_{\frac{1}{2},-\frac{1}{2}}(5)\eta_{\frac{1}{2},\frac{1}{2}}(6)
\right\},
\end{align}
where $\Phi_{0,i}^{(4)}$ is the $i$th basis
that gives the ground-state energy of $^4$He
with the full set of these basis functions.
The following two types of relative coordinates
are considered:\\
\noindent(i) Y-type
\begin{align}
\bm{y}_1^{({\rm Y})}=\bm{r}_5-\bm{x}_{\rm cm}^{(4)}, \quad
\bm{y}_2^{({\rm Y})}=\bm{r}_6-\frac{\bm{r}_5+4\bm{x}_{\rm cm}^{(4)}}{5},
\end{align}
(ii) T-type
\begin{align}
\bm{y}_1^{({\rm T})}=\bm{r}_6-\bm{r}_5, \quad
\bm{y}_2^{({\rm T})}=\frac{\bm{r}_5+\bm{r}_6}{2}-\bm{x}_{\rm cm}^{(4)}.
\end{align}
These configurations are essential for 
describing the two valence nucleon motion around the $\alpha$ core
which will be important, especially, in the low-lying energies.
For the Y-(T-)type channel, we assume that
both of the $\bm{y}_1^{({\rm X})}$ and $\bm{y}_2^{({\rm X})}$ coordinates
are initially in $p$($s$)-wave
and the one coordinate is excited to $s$($p$)-wave state.
We consider either $\bm{y}_1^{({\rm X})}$ or $\bm{y}_2^{({\rm X})}$ in
each coordinate set is excited by the $E1$ operator,
that is, the basis set with $l=1$ and 2 are independently included
respectively for (i) and (ii).
The relative wave functions
of the valence nucleons
are expanded with several Gaussian functions
covering from short to far distances, that is,
the diagonal matrix elements of a $2\times 2$ matrix $B$,
e.g., $B_{11}=1/b_{11}^2$, are chosen by
a geometric progression with 18 and 15 basis
ranging from 0.1 fm to 22 fm for the $\bm{y}_1^{({\rm X})}$
and $\bm{y}_2^{({\rm X})}$ coordinates, respectively.
For practical computations, we truncate the number of
the basis function of
the four nucleon subsystem, $\Phi_{0,i}^{(4)}$, with 15 basis.
Though the energy loss of this $\alpha$ particle is tiny in which
only 0.3 MeV difference from the full model space calculation is found,
it drastically reduces the total number of the basis functions.

The configurations of type (III) are defined as
\begin{align}
\Phi_{JM_J,ijk}^{(33)}&=
\mathcal{A}\left\{\left[\left[\Phi_{\frac{1}{2},i}^{(3)}(123)\Phi_{\frac{1}{2},j}^{(3)}(456)\right]_{J_3}\right.\right.\notag\\
&\times\left.\left.\exp\left(-\frac{1}{2}b_kz^2\right)
\mathcal{Y}_1(\bm{z})\right]_{JM_J}\right\}
\end{align}
with
\begin{align}
\bm{z}=\frac{\bm{r}_1+\bm{r}_2+\bm{r}_3}{3}
-\frac{\bm{r}_4+\bm{r}_5+\bm{r}_6}{3},
\end{align}
where $\Phi_{\frac{1}{2},i}^{(3)}(123)$ and $\Phi_{\frac{1}{2},j}^{(3)}(456)$
are the $i$th and $j$th bases
that give respectively the ground-state energies of $^3$He and $^3$H
with the full set of these basis functions.
These configurations describe the model space that directly excites
the $h+t$ cluster degrees-of-freedom imprinted on
the ground-wave function of $^6$Li ($S^2_{ht}\sim 0.4)$.
The relative wave function for the coordinate $\bm{z}$ is expanded
by 10 Gaussian functions with $p$-wave to reduce the computational cost. 
We also truncate the total number of basis functions
for the three-nucleon subsystems by 7 bases
resulting in only 0.2 MeV energy loss in these $h$ and $t$ particles.

Figure~\ref{channels.fig} shows schematic figures of 
the sets of the basis functions
explained above. Circles and thick line in red indicate the protons
and the coordinate excited by the $E1$ operator, respectively.
Note that we include all the basis states for each subsystem independently.
The final-state wave functions are not restricted
the subsystems being the ground state but
the excitations and distortion of $^6$Li, $\alpha$, $h$, and $t$ are
included through the coupling of the pseudo excited states of those
nuclear systems.
The number of bases included in this calculation
is 1800, 16200, and 490 for the configurations of types
(I), (II), and (III), respectively.
We diagonalize the Hamiltonian including all the configurations
with 18490 basis functions and find $\sim 2\times 10^3$ states below
the excitation energy of 100 MeV.

\section{Results and discussions}
\label{results.sec}

\subsection{Electric-dipole transitions and nuclear clustering}
\label{E1.sec}

\begin{figure}[ht]   
\begin{center}
\epsfig{file=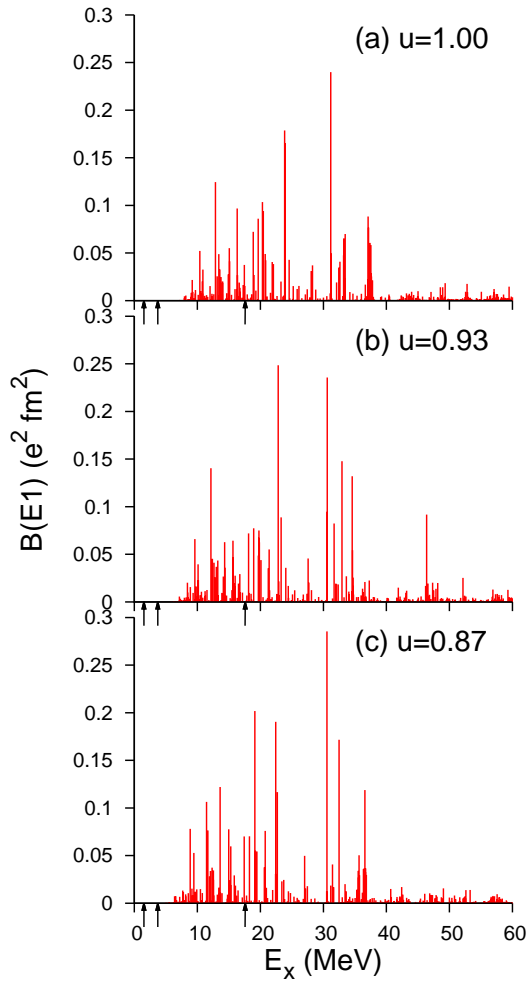, scale=1.1}    
\caption{Electric-dipole strengths of $^6$Li with the full model space
  with (a) $u=1.00$, (b) 0.93, and (c) 0.87.
 Arrows indicate the theoretical $\alpha+d$, $\alpha+p+n$, and $h+t$ thresholds
    from left to right, respectively.}
  \label{e1str.fig}
\end{center}
\end{figure}

\begin{figure}[ht]   
\begin{center}
\epsfig{file=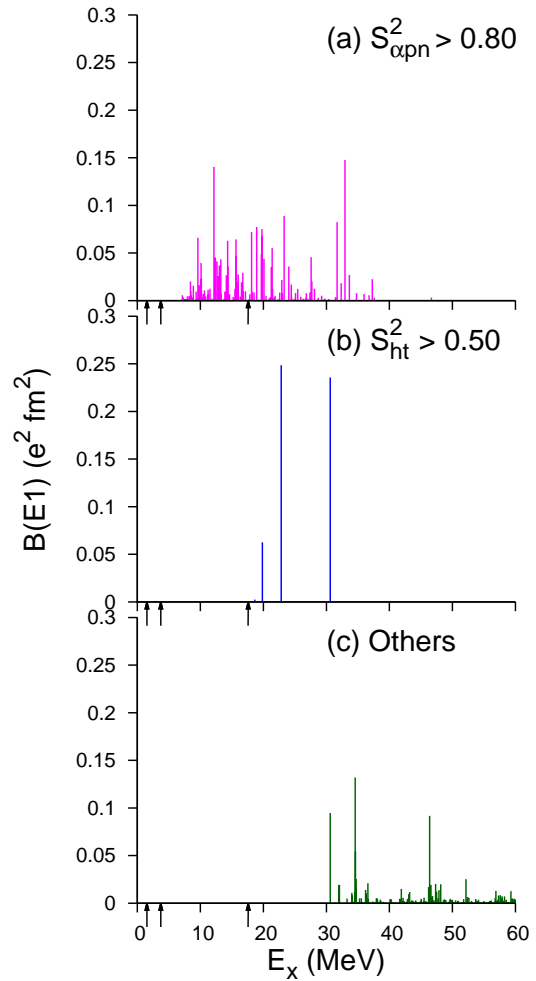, scale=1.1}    
\caption{Electric-dipole strengths of $^6$Li with the full model space
  with $u=0.93$ categorized by the amount of the spectroscopic factors
  into three:
  (a) $S_{\alpha pn}^2 >0.80$, (b) $S_{ht}^2 >0.5$, and (c)
  neither (a) nor (b).   See text for details.
  Arrows indicate the theoretical $\alpha+d$, $\alpha+p+n$,
  and $h+t$ thresholds
   from left to right, respectively.}
  \label{e1str_decom.fig}
\end{center}
\end{figure}

Figure~\ref{e1str.fig} plots
the $E1$ transition strengths
obtained with the full model space
that includes the configurations of types (I)--(III) described
in the previous section with different values of the
$u$ parameter as a function of the excitation energy, $E_x=E_f-E_0$.
For all these $u$ values,
we see several large $B(E1)$ values in the low ($E_x\lesssim 16$ MeV),
intermediate ($E_x\sim 16$--30 MeV), and high ($E_x\gtrsim 30$ MeV)
energy regions.
Though there is little quantitative difference among
these three different cases up to $E_x\sim$40 MeV,
hereafter we discuss the results with $u=0.93$ unless otherwise
mentioned.

The structure of those peaks becomes more transparent by categorizing those
states with respect to the spectroscopic factors of the $\alpha+p+n$
configuration
\begin{align}
  &S_{\alpha pn}^2=\left.\left|\left<\Psi^{(\alpha)}\Psi^{(p)}\Psi^{(n)}\right.|\Psi^{(6)}_{JM_J}(E)\right>\right|^2,
\label{SFapn.eq}
\end{align}
where $\Psi^{(\alpha)}$
is the ground-state wave function of $^4$He,
and $\Psi^{(p)}$ ($\Psi^{(n)}$) is the proton (neutron) wave function.
All the relative coordinates between clusters,
spins and isospins are integrated out.
Details about the evaluation are given in Appendix.
Note that $S^2_{\alpha d}$ is a subset of $S_{\alpha pn}^2$ but
the states with large $S^2_{\alpha d}$ value
does not contribute to the $E1$ transition.
No $E1$ transition occurs from the ground state
to those states because their total isospins are almost 0.

Panel (a) of Fig.~\ref{e1str_decom.fig} displays
the transition strengths to the states with
$S_{\alpha pn}^2 >0.80$. We find that
the $E1$ transition strengths distribute in ranges from 10 to 40 MeV
and that most of the low-lying states below 20 MeV have
a large $\alpha+p+n$ cluster component, which is consistent
by reminding the facts that the lowest $h+t$
threshold is 15.8 MeV~\cite{Tilley02}.

Panel (b) of Fig.~\ref{e1str_decom.fig} shows the $E1$ strengths
with $S_{ht}^2>0.50$. Three large $E1$ strengths
appear after the $h+t$ threshold opens.
We find that these peak structures are robust, and
their positions and strengths do not depend much
on the values of the $u$ parameter.
The first two structures may correspond to the observed $J^\pi=2^-$
levels at $E_x=17.98$ and 26.59 MeV with the $h+t$ decay widths
of 3.0 and 8.7 MeV, respectively~\cite{Tilley02}.

The other strengths which cannot be categorized
into the above two conditions are plotted in the panel (c)
of Fig.~\ref{e1str_decom.fig}.
They appear beyond 30 MeV and some prominent $E1$ strengths are found
between 30 and 40 MeV. In this energy region,
the all particle thresholds are open.
Various configurations can couple with each other.
We will discuss more details in Sec.~\ref{trdens.sec}.

\begin{table}[htb]
\begin{center}
  \caption{Excitation energy ($E_x$), reduced electric-dipole
    transition probability [$B(E1)$] in unit of $e^2$fm$^2$, and $\alpha+p+n$ ($S_{\alpha pn}^2$)
    and $h+t$ ($S_{ht}^2$) spectroscopic factors of $^6$Li
    with $u=0.93$. Note that two prominent strengths
    with different configurations appear at
    $E_x=30.6$ MeV.}
    \begin{tabular}{cccc}
      \hline\hline
      $E_x$(MeV)&$B(E1)$&$S_{\alpha pn}^2$&$S_{ht}^2$\\
\hline
 9.6&0.066&0.999&0.000\\
12.1&0.140&0.999&0.011\\
14.3&0.063&0.997&0.000\\
15.7&0.064&0.999&0.010\\
\hline
18.9&0.077&0.991&0.004\\
19.8&0.075&0.994&0.003\\
22.8&0.249&0.113&0.850\\
23.3&0.089&0.963&0.003\\
\hline
30.6&0.236&0.264&0.533\\
30.6&0.095&0.780&0.158\\
33.0&0.148&0.962&0.005\\
34.6&0.132&0.195&0.019\\
\hline\hline
  \end{tabular}
  \label{spect.tab}
\end{center}
\end{table}

We discuss the structure of the states with large $E1$ strengths.
For quantitative discussions, we list, in Table~\ref{spect.tab},
$E_x$, $B(E1)$, $S_{\alpha pn}^2$, and $S_{ht}^2$
of the states giving four largest $B(E1)$ values
in the low-, intermediate-, and high-energy regions with $u=0.93$.
In the low energy regions ($E_x\lesssim$16 MeV) below
the $h+t$ threshold, all the states have large
$S_{\alpha pn}^2$ values being almost unity,
whereas their $S_{ht}^2$ values are almost zero.

In the intermediate energy regions around 20 MeV where
$\alpha$-cluster can be energetically possible to break,
although most of the states still have large $S_{\alpha pn}^2$ values,
the state at $E_x=22.8$ MeV have a small $S_{\alpha pn}^2$ value
in which the $\alpha$ cluster in the six-nucleon
system is strongly distorted.
This state is dominated by the $h+t$ configuration
having the large $S_{ht}^2$ value 0.850.

In the high energy regions beyond $\sim$30 MeV
where all the particle thresholds are open,
various structures are found: Mixture of $\alpha+p+n$
and $h+t$ components at $E_x=30.6$ MeV,
almost pure $\alpha+p+n$ component at $E_x=33.0$ MeV,
and neither $\alpha+p+n$ nor $h+t$ components at $E_x=34.6$ MeV.

\begin{figure}[ht]   
\begin{center}
\epsfig{file=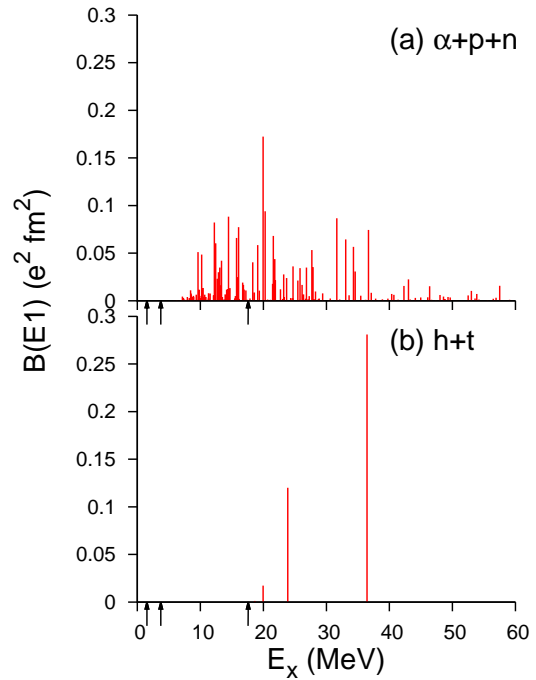, scale=1.1}    
  \caption{Electric-dipole strengths only with
    (a) $\alpha+p+n$ and (b) $h+t$ final-state configurations.
    Arrows indicate the theoretical $\alpha+d$, $\alpha+p+n$,
    and $h+t$ thresholds
    from left to right, respectively.}
  \label{e1str-part.fig}
\end{center}
\end{figure}

We have shown the $E1$ strength distributions with the full model space
including the breaking and polarization of the clusters in $^6$Li. 
To make the role of these effects clearer,
we calculate the $E1$ transition strengths only with the $\alpha+p+n$
and $h+t$ configurations that are respectively constructed by
the diagonalization of the following basis functions
\begin{align}
  \Phi_{JM_J,jk}^{(\alpha pn),m}&=\sum_{i}C_i^{(\alpha)}\Phi_{JM_J,ijk}^{411,m},
  \label{apnconf.eq}\\
  \Phi_{JM_J,k}^{(ht)}&=\sum_{i,j}C_i^{(h)}C_j^{(t)}\Phi_{JM_J,ijk}^{(33)},
\label{htconf.eq} 
\end{align}
where $C_i^{(x)}$ are a set of the coefficients 
that give the ground-state wave function of a cluster
$x(= h$, $t$,  and $\alpha$). 
Figure~\ref{e1str-part.fig} plots the transition strengths only 
with the $\alpha+p+n$ and $h+t$ final-state configurations.
As expected the transition strengths only with the $\alpha+p+n$ configurations
below 30 MeV are similar to the results
displayed in the panel (a) of Fig.~\ref{e1str_decom.fig}
where the $S_{\alpha pn}^2$ values are large.
We only find three significant strengths with the $h+t$ configurations
and the positions of the two lowest prominent peaks at $E_x=20.0$ and 23.9 MeV
remain unchanged with the full model space calculations,
while the most prominent peak at $E_x=36.4$ MeV split into small
strengths in the full model space calculation as shown
in the panel (b) of Fig.~\ref{e1str.fig}.
We find the state has relatively large square overlap
with the $\alpha+p+n$ configuration (0.281),
leading to the level splitting due to the channel coupling.

\subsection{Non-energy weighted sumrule}

\begin{figure}[ht]   
\begin{center}
\epsfig{file=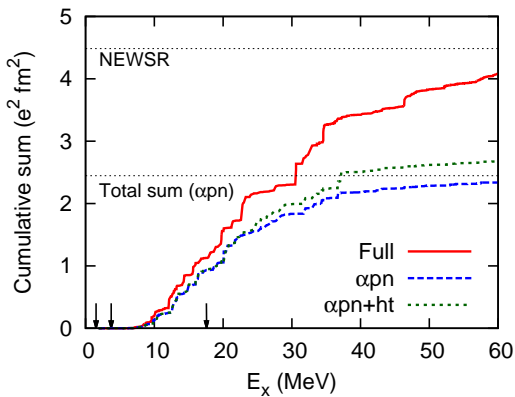, scale=1.2}    
\caption{Cumulative sum of the $E1$ transition strengths of $^6$Li
  with the full model space, $\alpha+p+n$,
  and $(\alpha+p+n)+(h+t)$ configurations with $u=0.93$.
  The non-energy-weighted sumrule (NEWSR) value and
  the total sum of the $E1$ strengths only with the $\alpha+p+n$
  configurations are plotted as thin dotted lines for comparison.
  See text for details. 
Arrows indicate the theoretical $\alpha+d$, $\alpha+p+n$ and $h+t$ thresholds
    from left to right, respectively.
}
  \label{e1sum.fig}
\end{center}
\end{figure}

Here we discuss the impact of the clustering configurations
on the $E1$ sumrule.
The non-energy-weighted sumrule (NEWSR) can be evaluated by
\begin{align}
  \sum_{E_f}B(E1,E_f)=e^2\left(Z^2r_p^2-
  \frac{Z(Z-1)}{2}r_{pp}^2\right).
\label{sumrule.eq}
\end{align}
We obtain 4.49 $e^2$fm$^2$ as a total sum of
the $E1$ transition strengths with the full model space.
The NEWSR is fully satisfied, that is,
99.6\% of the right-hand-side of Eq.~(\ref{sumrule.eq})
is fulfilled in the present model space.
To quantify the importance of the $\alpha$ clustering,
we calculate the left-hand-side equation
only with the $\alpha+p+n$ configuration defined in Eq. (\ref{apnconf.eq})
and the value is 2.45 $e^2$fm$^2$, satisfying 55\% of the total sumrule value.
Figure~\ref{e1sum.fig} compares the cumulative sum
of $^6$Li of the $E1$ strengths with the full model space
as well as the ones only with the $\alpha+p+n$ configurations.
To get the sumrule satisfied, say 80\%,
the cumulative sum with the full model space
needs to integrate up to about 45 MeV, whereas
the most of the important configurations
with the $\alpha+p+n$ configuration are exhausted
at 33 MeV where its cumulative sum exceeds 80\% of its total sum.
We find that the configurations other than
the $\alpha+p+n$ configurations are also important
in such low-lying regions below $\sim$20 MeV
even the $h+t$ threshold does not open.
They are used to describe the polarization of the clusters
through the coupling of those cluster configurations.
The difference between the cumulative sum of
the full model space and $\alpha+p+n$ are 30\% at 10 MeV
and the difference becomes large as the incident energy increases.
We also display in Fig.~\ref{e1sum.fig}
the cumulative sum of the transition strengths with
the mixing of $\alpha+p+n$ and $h+t$ configurations defined
respectively in Eqs. (\ref{apnconf.eq}) and (\ref{htconf.eq}).
The $h+t$ configurations play a role beyond $\sim 20$ MeV
after opening the $h+t$ threshold and improve the NEWSR value by 8\%.
However, it is not enough to explain all the needed configurations
included in the strengths with the full model space.
In the higher energies,
the breaking of these cluster configurations becomes more important
as various configurations can contribute to the $E1$ transitions.

\subsection{Structure of the electric-dipole excitation}
\label{trdens.sec}

\begin{figure}[ht]   
\begin{center}
\epsfig{file=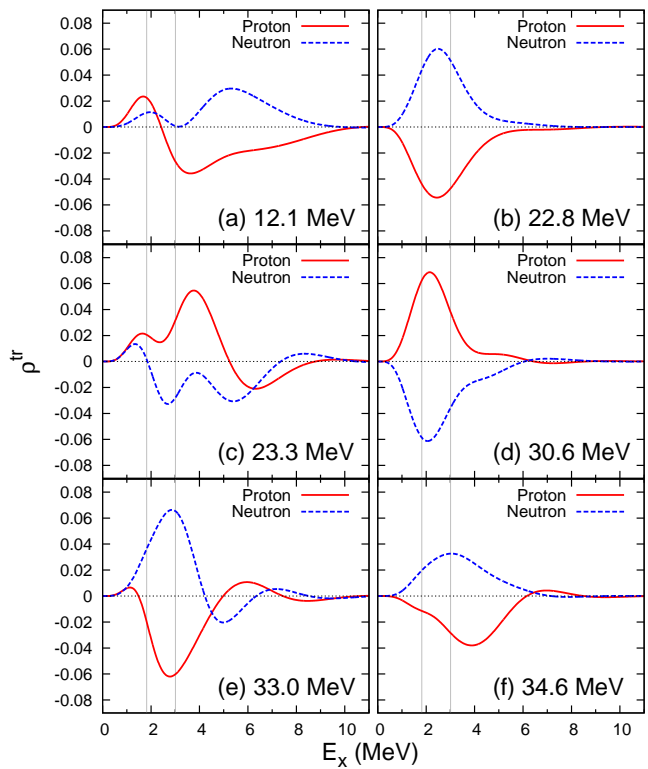, scale=0.9}    
  \caption{Transition densities of $^6$Li with the excitation energy 
    of (a) 12.1, (b) 22.8, (c) 23.3, (d) 30.6, (e) 33.0, and (f) 34.6 MeV
    selected from Table~\ref{spect.tab}. See text for details.
    Note that left panels
    plot the transition densities with the states having large
    $S_{\alpha pn}^2\sim 1$.
    Vertical thin lines indicate
    theoretical nuclear radii, $\sqrt{\frac{5}{3}}r_m$,
    with $r_m=1.41$ and 2.33 fm for $^4$He and $^6$Li, respectively.}
  \label{trdens_full.fig}
\end{center}
\end{figure}

Let us discuss the spatial structure of the $E1$ transitions.
For this purpose,
we calculate the transition densities of $^6$Li
for the analysis of the $E1$ transition mode~\cite{Mikami14}
\begin{align}
\rho^{\rm tr}_{p/n}(r)=
\sum_{i\in p/n}\left<\Psi_{J_f}^{(6)}\right\|
\mathcal{Y}_{1}(\bm{r}_i-\bm{x}_{6})\delta(|\bm{r}_i-\bm{x}_6|-r)
\left\|\Psi_{J_0}^{(6)}\right>,
\end{align}
for proton and neutron.
This quantities represent the spatial distributions
of the proton and neutron transition matrices reminding that
the $E1$ transition matrix can be obtained with
\begin{align}
\left<\Psi_{J_f}^{(6)}\right\|\mathcal{M}(E1)\left\|\Psi_{J_0}^{(6)}\right>=
e\sqrt\frac{4\pi}{3}\int_0^\infty dr\,\rho^{\rm tr}_{p}(r).
\end{align}

We discuss the transition densities
of $^{6}$Li to the selected states that show some characteristic behaviors.
Figure~\ref{trdens_full.fig}
plots the transition densities for proton and neutron that
correspond to the prominent $B(E1)$ peaks with $B(E1) > 0.1$ $e^2$fm$^2$.
At the low energy (a) $E_x=12.1$ MeV, we see the in-phase transition
below the $^6$Li radius and the out-of-phase transitions
occur outside the nuclear surface.
This characteristic transition can be interpreted as the GDR-like
or Goldhaber-Teller-(GT)-dipole type oscillation~\cite{Goldhaber48}
of the valence proton and neutron around the core ($\alpha$).
In fact, the state have large $S_{\alpha pn}^2$ value as listed in
Table~\ref{spect.tab}.
This type of ``soft'' GT-dipole mode is very unique in $^6$Li
and differs from the soft-dipole mode
expected in $^6$He that the oscillation between
the valence two neutrons against the core.
We note that the $E1$ transitions to $\alpha+d$ 
is almost forbidden because the $E1$ operator is isovector
and the ground state of $^6$Li is almost pure $T=0$ state though
small mixture of the other total isospin state is included.
Actually, the $S_{\alpha d}^2$ value for this state is 0.02. 
Studying the low-lying $E1$ excitations of $^6$Li is
the ideal example that the soft GT-dipole mode dominates.

In the energy regions where the $h+t$ threshold opens,
we see clear out-of-phase transitions in all regions
at (b) $E_x=22.8$\,MeV which is typical for the GDR mode.
As $S_{ht}^2$ is large $>0.8$,
this behavior comes from the $E1$ excitation of
the relative motion between the $h$ and $t$ clusters.
Peak positions are at $\sim$2 fm
located at the sum of the peak positions
of the density distributions of $^3$H and $^3$He displayed in
Fig.~\ref{dens.fig}.
Such cluster GT-dipole modes can appear in any nuclear systems
but its emergence depends on the location of the cluster threshold.
In light nuclei, since the cluster threshold becomes low,
the cluster GT dipole modes can appear in the lower-lying regions.
For the $^{7}$Li case, the $\alpha+t$ threshold
is lowest (2.47 MeV~\cite{Tilley02}),
it differs from the case of $^6$Li,
the cluster GT-dipole mode is expected to appear
in the lowest energy regions.

We also find large $B(E1)$ value at almost the same energy
(c) $E_x=23.3$ MeV. Similarly to the transition density at
(a) $E_x=12.1$ MeV,
we see the in-phase transition below the $^{6}$Li radius
but more oscillations in the out-of-phase transition appear in
the outside of the nuclear surface.
Since this state has large $S_{\alpha pn}^2$ value
listed in Table~\ref{spect.tab},
this state can be interpreted as a vibrational excitation
of the soft GT-dipole mode. The transition densities
with (e) $E_x=33.0$ shows the similar character having more oscillations.

At (d) $E_x=30.6$ MeV, this shows the out-of-phase transition
in all regions. Since the peak position is
almost the same as that of (b) $E_x=23.3$ MeV
and has large mixture of $h+t$ configurations $\sim 0.5$
listed in Table~\ref{spect.tab}, this state
can be regarded as a vibrational excitation of
of the state with (c) $E_x=23.3$ MeV
which exhibits the $h+t$ clustering.

Finally, the state with (f) $E_x=34.6$ MeV
shows also the out-of-phase transition
in all regions but the peak positions are located outside
of the nuclear surface showing totally different structure from
that of the $h+t$ oscillation.
Neither $S_{\alpha pn}^2$ nor $S_{ht}^2$ is large
as listed in Table~\ref{spect.tab}.
This state can be regarded as having the typical GDR structure
that the protons and neutrons
oscillate opposite to each other~\cite{Goldhaber48}.

\begin{figure}[ht]   
\begin{center}
\epsfig{file=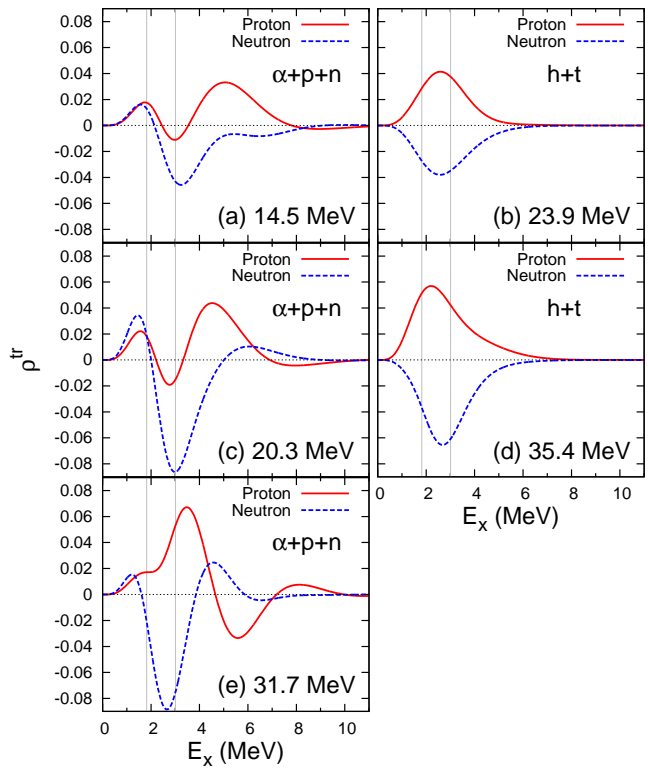, scale=0.9}    
  \caption{Transition densities to the states
    that give the prominent $E1$ strength only with the restricted
    model spaces: The $\alpha+p+n$ final-state configurations
    at (a) $E_x=14.5$, (c) 20.3, and (e) 31.7 MeV;
    and the $h+t$ final-state configurations at (b) $E_x=23.9$
    and (d) 35.4 MeV. See text for details.
    Vertical thin lines indicate
    theoretical nuclear radii, $\sqrt\frac{5}{3}r_m$,
    of $^4$He and $^6$Li, respectively.}
  \label{trdens-part.fig}
\end{center}
\end{figure}

To strengthen the interpretations given above,
we calculate the transition densities with the final-state
wave functions only with the $\alpha+p+n$ and $h+t$ configurations.
These transition strengths with the limited model spaces
were already given in Fig.~\ref{e1str-part.fig}.
The transition densities with the $\alpha+p+n$ configuration
that give the three largest
$B(E1)$ strengths for each configuration
are shown in the panels (a), (c), and (e) of Fig.~\ref{trdens-part.fig}.
These almost explain the characteristics behavior
of the transition densities of the states with large $S_{\alpha pn}^2$,
that are, the soft GT-dipole modes.
All the transition densities
have in-phase transitions inside about the nuclear radius 
and out-of-phase transitions beyond the surface.
It is interesting to note that
the nodal or oscillatory behavior in the in-phase regions
of the transition densities at around the radius of the $\alpha$ particle.
This is due to the Pauli principle between the core and
valence nucleon. 
As we see in the transition densities of $^6$He,
the nodal behavior of the transition density
can only be seen in the neutron transition~\cite{Mikami14}.

We also plot, in the panels (b) and (d) of Fig.~\ref{trdens-part.fig},
the transition densities only with the $h+t$
configurations for the states giving the two highest $E1$ strengths.
They show typical GDR behavior and peak positions
are at around the nuclear surface which explains
the behavior of the transition densities
of (b) and (d) of Fig.~\ref{trdens_full.fig}.
In this restricted model space,
we do not obtain the similar transition densities to the state
with $E_x=34.6$ MeV, implying that these clusters
are strongly distorted
as was shown in the GDR mode
at $E_x=32.9$ MeV in $^6$He~\cite{Mikami14}.

In summary, various types of the $E1$ excitations of $^6$Li
can be classified by focusing on the nuclear clustering.
The panels (a), (c), and (e) of Fig.~\ref{trdens_full.fig}
have the same characteristics
that the in-phase transition below the $^6$Li radius due
to the $\alpha$ clustering and out-of-phase transitions of the valence nucleons
beyond the nuclear surface (Soft GT-dipole mode).
The figures (b), (d), and (f) of Fig.~\ref{trdens_full.fig}
show the out-of-phase transition in all regions:
The excitation modes of (b) and (d)
are originated from the oscillations between $h$ and $t$ clusters
(Cluster GT-dipole mode),
and (f) is the typical GDR oscillation that protons and neutrons
oscillate opposite to each other (GT-dipole mode).

\subsection{Photoabsorption cross sections}

The total photoabsorption cross section
is calculated by using the formula~\cite{RS}
\begin{align}
  \sigma_\gamma(E_\gamma)= \frac{4\pi^2}{\hbar c}E_\gamma\frac{1}{3}
  \frac{dB(E1,E_\gamma)}{dE},
\end{align}
The continuum states are discretized in this calculation.
For a practical reason, they are often smeared by the Lorentzian functions
$dB(E1,E)/dE=\frac{\Gamma}{2\pi}\sum_\nu B(E1,E_\nu)/[(E-E_\nu-E_0)^2+(\Gamma/2)^2]$, using a certain $\Gamma$ value as a free parameter~\cite{Pinilla11}.
To compare with the recent experiment,
we fix the width parameter so as to reproduce the 
total sum of the experimental cross sections of Ref.~\cite{Yamagata17}.
However, the energy-independent width does not work
that results in the unphysically large decay width $\sim$40 MeV.
Thus, we use the energy dependent decay width that starts from
the the lowest threshold: $\Gamma(E)=[E-E(\alpha+d)]\tan 2\theta$,
where $E(\alpha+d)$ is the $\alpha+d$ threshold energy.
The $\theta$ value is determined to $27^\circ$ which reproduces
the the total sum of the cross sections of Ref.~\cite{Yamagata17}.

\begin{figure}[ht]   
\begin{center}
\epsfig{file=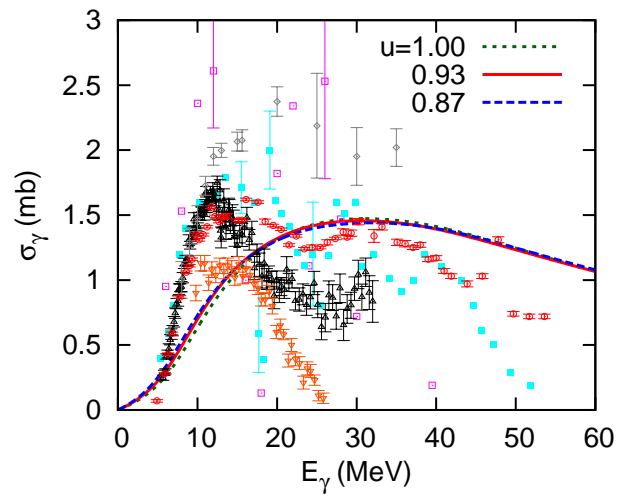, scale=1.5}    
\caption{Comparison of the photoabsorption cross sections of $^6$Li.
  Experimental $^6$Li$(\gamma,n)$ data are plotted
  as open rectangles~\cite{Costa63}, closed rectangles~\cite{Bazhanov65},
  open triangles~\cite{Berman65}, inverted open triangles~\cite{Denisov67},
  and diamonds~\cite{Wurtz14}. Since they are unavailable,
  most of error bars of the experimental data
  in Refs.~\cite{Costa63, Denisov67} are not plotted.
  Open circles stand for
  the $^{6}$Li$(\gamma,n)+^{6}$Li$(\gamma,2n)+^{6}$Li$(\gamma,3n)$
    data taken from Ref.\cite{Yamagata17}.}
  \label{photo.fig}
\end{center}
\end{figure}

Figure~\ref{photo.fig} compares the calculated 
and experimental photoabsorption cross sections
of $^6$Li. Most of the experimental data are
the cross sections for the $^{6}$Li($\gamma,n$) reactions
but the $^{6}$Li($\gamma,2n$) and
$^{6}$Li($\gamma,3n$) cross sections are negligibly small
in this energy region~\cite{Yamagata17}.
In the calculated total photoabsorption cross sections,
only a single-peak structure is found.
We remark that Ref.~\cite{Bacca02, Bacca04}
also predicted a broad single-peak structure at around 20 MeV
for the total photoabsorption cross sections of $^6$Li
with the six-body calculation.
The results are almost identical for all $u$ parameters
due to large smearing width $\Gamma(E)$.
Though all fine structures are smeared out, 
the cross section values are quantitatively consistent
with the measured cross sections
considering that the measured values are very much scattered.
For the quantitative comparison to the measured cross sections,
it is necessary to describe the six-body 
continuum states appropriately with the aid of, e.g., the complex
scaling method~\cite{Moiseyev98,Aoyama06} as well as the improvement of
the nuclear interaction, although they are involved.

Let us compare the interpretation
given in Refs.~\cite{Costa63,Yamagata17} and our findings.
The measured cross sections of Ref.~\cite{Yamagata17} show
the two-peak structure and their interpretation on the two-peak structure
was that, the low-lying ($E_\gamma\lesssim 20$ MeV) peak comes from
the typical GDR transition mode
of $^6$Li and the higher peak (from $\sim 30$ to $\sim 40$ MeV)
corresponds to the GDR of the $\alpha$ cluster
in $^6$Li. Contrary to that interpretation, 
we see the typical GDR or GT mode appears
in the higher-lying energy regions around 35 MeV,
where the $\alpha$ cluster is strongly distorted.
In Ref.~\cite{Costa63}, the lower peak is interpreted as
the disintegration to the $\alpha+p+n$ channels, whereas
the higher peak is the GDR of $^6$Li
due to the disintegration of the $\alpha$ core. 
From the theoretical point of view, it is difficult to say
whether this GDR mode is the GDR of the $\alpha$ cluster in $^6$Li or not
in the high energy regions
because identical fermions cannot be distinguished where they belong to.
In our interpretation, in the low-lying energy regions below the $h+t$
threshold (15.8 MeV), the soft GT-dipole transitions dominate
that the out-of-phase transition between
the valence nucleons around the $\alpha$ cluster,
which is consistent with the interpretation
given in Ref.~\cite{Costa63} for the lower energy peak.
This can be realized as all the spectroscopic factors, $S_{\alpha pn}^2$,
of the final states in this energy regions are almost unity.
It is known that the excitation energy of the GDR
is inversely proportional to the nuclear radius.
According to the systematics of the GDR energy~\cite{RS},
this low-lying energy region corresponds
to the GDR energies of $A\sim 200$ nuclei.
Therefore, it is natural to interpret
that the typical excitation mode in this energy region
is the GT mode of the valence nucleons around
the tightly bound $\alpha$ core, whereas
the typical GDR mode of $^{6}$Li suggested
in Ref.~\cite{Yamagata17} is unlikely,
reminding that the $^6$Li radius is about one half
of the radii of the $A\sim 200$ nuclei~\cite{Angeli13}.
In the intermediate energies from $\sim 20$ to $\sim 30$ MeV
just between the low and high-lying peaks of $^{6}$Li,
the prominent $h+t$ cluster GT-dipole mode appears,
which is consistent with the interpretation
given in Ref.~\cite{Costa63} for the higher energy peak.
As summarized at the end of the previous subsection,
the emergence of these various excitation modes
can simply be recognized by the threshold energies,
the Ikeda threshold rule~\cite{Ikeda68}.

\subsection{Isoscalar dipole transitions}

Here we discuss an other operator to discuss more details on
the transition densities.
The compressive isoscalar dipole (IS1) 
operator~\cite{Stringari82} is defined by
\begin{align}
  \mathcal{M}({\rm IS}1)
  =\sum_i(\bm{r}_i-\bm{x}_N)^2\mathcal{Y}_{1\mu}(\bm{r}_i-\bm{x}_N).
\end{align}
The transition matrix of IS1 
can be calculated by using the relation:
$\int_{0}^\infty dr\, r^2 \left(\rho_{p}^{\rm tr}+\rho_{n}^{\rm tr}\right)$.
The IS1 transitions have recently been intensively discussed
because it has of particular importance
to study the cluster structure (See recent theoretical
and experimental papers~\cite{Chiba17,Enyo17,Adachi18}
and references therein).

\begin{figure}[ht]   
\begin{center}
\epsfig{file=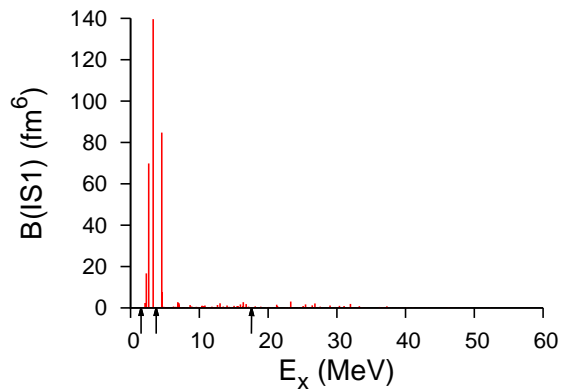, scale=1.2}    
\caption{Isoscalar dipole strengths of $^6$Li
  as a function of the excitation energy.
Arrows indicate the theoretical $\alpha+d$, $\alpha+p+n$ and $h+t$ thresholds
    from left to right, respectively.}
  \label{dipolestr.fig}
\end{center}
\end{figure}

Figure~\ref{dipolestr.fig} plots the IS1 strength distributions
as a function of the excitation energies. 
We see some prominent strengths below 5 MeV 
having the isoscalar nature possibly by the $\alpha+d$ continuum, 
which cannot be excited by the $E1$ operator that only has
the isovector term. 
In fact, the $S_{\alpha d}^2$ values of those states
are found to be almost unity
and the transition densities of the state with 
the most prominent IS1 peak at $E_x=3.3$ MeV shows
in-phase transition in all regions.
Beyond 5 MeV, the IS1 strengths
drop suddenly and very small strengths appear in the higher energy regions.
The most of the IS1 strengths are exhausted
by the transitions to the $\alpha+d$ states below 5 MeV.
In case of $^6$He~\cite{Mikami14},
since the low-lying soft-dipole mode is dominated
by the surface excitation of the valence neutrons,
several IS1 strengths appear at the low-lying regions. 
Contrary to the $^6$He case,
the states with all the prominent $E1$ strengths
have the out-of-phase excitation character.
The IS1 transition matrix which is obtained by the sum
of proton and neutron transition densities is strongly canceled out
in such excitation modes where the out-of-phase
transitions dominate.
The contributions from the in-phase transition regions
in the soft GT-dipole mode become small
in the IS1 strengths due to the additional $r^2$ factor
appearing in the IS1 operator.
The strong suppression of the IS1 transition strengths
can be the evidence that all the $^6$Li final states
beyond 5 MeV are dominated by the out-of-phase transitions.
We remark that similar transition strengths are observed
in the proton inelastic scattering on $^{6}$Li~\cite{Yamagata06},
although a proton probe can excite both the isoscalar and isovector
components.
The experimental confirmation using an isoscalar probe
such as an $\alpha$ particle is desired to
clarify the excitation mechanism of $^6$Li.

\section{Conclusion}
\label{conclusion.sec}

Motivated by the recent measurement of the photoabsorption cross sections
of $^6$Li~\cite{Yamagata17},
we have performed fully microscopic six-body calculations
for the electric-dipole ($E1$) transition strengths.
The ground-state wave function of $^6$Li was obtained precisely
by using the correlated Gaussian (CG) functions with the stochastic
variational method.
The final-state wave functions populated by the $E1$ operator
were expanded by a number of the CG functions
including the explicit asymptotic cluster
wave functions as well as their distorted configurations
that are important to describe the complicated six-nucleon dynamics.
Emergence of various excitation modes has been found
through the analysis of the transition densities from the ground
to final-state wave functions.
The degrees of the clustering in those states have been quantified
by evaluating the components of the $\alpha+p+n$ and $h+t$ clusters
in the $^6$Li wave functions to understand
the role of these cluster configurations in the $E1$ excitations.

Nuclear clustering plays a crucial role in explaining the $E1$ 
excitation mechanism of $^6$Li and its emergence 
strongly depends on the positions of the threshold energies.
In the low energy regions below the $\alpha$ 
breaking or $h+t$ threshold energy $\lesssim 16$ MeV,
we found that the $E1$ excitations are dominated by
the ``soft'' dipole mode that
exhibits the in-phase transitions of proton and neutron transition
densities in the internal regions
and the out-of-phase transitions beyond the nuclear surface.
This can be interpreted as the out-of-phase oscillation between
valence nucleons around the $\alpha$ cluster in $^6$Li
[``soft'' Goldhaber-Teller(GT)-dipole mode],
which is a very unique excitation mode.
After the $h+t$ thresholds open,
the $h+t$ cluster mode appears
showing the out-of-phase transition in all regions
and they also compete with the vibrational excitation
of the soft GT-dipole modes having the $\alpha+p+n$ structure.
Beyond 30 MeV, where all decay channels open,
$\alpha+p+n$ and $h+t$ and other possible channels
can mix and compete and finally typical GDR mode appear in this
energy regions.

These interpretations
are different from the speculation given in Ref.~\cite{Yamagata17}
that the low-energy peak corresponds to the typical GDR of $^6$Li.
From the present analysis, we found that
the $E1$ transition strengths of the $^6$Li are dominated by the
out-of-phase transitions of protons and neutrons in the surface regions
from the low- to high-energy regions, which is in contrast
to $^6$He where the neutron transition dominates at the low energy regions.
This phase property can be verified by using an isoscaler probe such as
$\alpha$ inelastic scattering measurement to confirm whether no prominent
strength is found or not after the $\alpha+p+n$ threshold. 

It is interesting to explore whether the soft GT mode appears
in the low-lying energy regions of heavier nuclei.
Since the excitation mode emerges
from the out-of-phase transition of the proton and neutron of
the $d$-cluster around the core in the initial ground-state wave function,
the ground-state wave function should have
a well-developed core plus $d$-cluster structure.
The most probable candidate is $^{18}$F because
the ground-state spin-parity is $1^+$ like $^6$Li
and a $^{16}$O$+p+n$ cluster structure component can be large.

Also, as a natural extension of $^6$Li,
a nucleus $^7$Li is worth studying~\cite{Yamagata17}.
Since the $\alpha+t$ threshold is the lowest (2.47 MeV),
the cluster GT mode of $\alpha+t$ is expected to appear first,
and then the other excitation modes appear with respect to
the opening of the particle decay channels
$\alpha+d+p$, $\alpha+p+p+n$, $t+h+n$, etc., in order.

These studies will serve as the universal understanding of
the emergence of the nuclear clustering
and reveal the excitation mechanism of nuclei through
the $E1$ field, which is one of the most simplest probes
of the nuclear structure.

\acknowledgments

The authors thank J. Singh for a careful reading of the manuscript.
This work was in part supported by JSPS KAKENHI Grant Numbers
18K03635, 18H04569, and 19H05140, and
the collaborative research program 2018,
information initiative center, Hokkaido University.

\appendix

\section{Calculation of the spectroscopic factors}

As a measure of degrees of the clustering, we evaluate
the spectroscopic factors, that are, the components of finding 
the $\alpha$, $\alpha+d$, and $h+t$ configurations
in the wave function of $^6$Li.
Eqs. (\ref{SFapn.eq}) and (\ref{SFab.eq}),
are respectively written more explicitly as
\begin{align}
  S_{\alpha pn}^2
  &=\biggl|\iint d\bm{r}\,d\bm{r}^\prime
  \left<\Psi^{(\alpha)}\Psi^{(p)}\Psi^{(n)}\right|\biggr.\notag\\
  &\biggl.\times\delta(\bm{y}_1-\bm{r})\delta(\bm{y}_2-\bm{r}^\prime)
  \left|\Psi^{(6)}_{JM_J}(E)\right>\biggr|^2,\\
  S_{ht}^2&=\left|\int  d\bm{r}
  \left<\left.\Psi^{(h)}\Psi^{(t)}\delta(\bm{z}-\bm{r})
  \right|\Psi^{(6)}_{JM_J}(E)\right>\right|^2.
\end{align}
All the relative wave functions are integrated out by using 
the orthonormal basis $\sum_{lm}\phi_{l,i}(r)Y_{lm}(\hat{\bm{r}})$
constructed from a sufficient number of
Gaussian functions, $r^l\exp(-ar^2)$, as a complete set.
Practically, we make the orthonormal basis sets by diagonalizing
the relative wave functions used
in the final-state wave function of types (II) and (III).
More explicitly, 
we diagonalize the following overlap matrices
using the coefficients of the bases
that give the ground-state wave function of a nucleus $x$, $C_i^{(x)}$
for the $\alpha+p+n$ spectroscopic factor 
\begin{align}
B_{mn}^{(\alpha pn)}=\sum_{i,j}^{K_4}C_i^{(\alpha)}C_j^{(\alpha)}\left<\Phi_{JM_J,im}^{(411,k)}\left|\Phi_{JM_J,jn}^{(411,l)}\right.\right>
\end{align}
with $k$ and $l$ run for 1 and 2 corresponding to
the Y- and T-types, respectively.
We take $K_4=15$. In the end, the dimension of $B^{(\alpha pn)}$ is 1080.
For the $h+t$ spectroscopic factors, we diagonalize the following
overlap matrix
\begin{align}
B_{mn}^{(ht)}=\sum_{i,j,k,l}^{K_3}C_i^{(h)}C_j^{(t)}C_k^{(h)}C_l^{(t)}\left<\Phi_{JM_J,ijm}^{(33)}\left|\Phi_{JM_J,kln}^{(33)}\right.\right>.
\end{align}
Finally, all the spectroscopic factors calculated in this paper
are evaluated by the overlap matrix element
of the correlated Gaussians~\cite{SVM,Suzuki08}.
The same procedure is applied for
the evaluation of the $\alpha+d$ spectroscopic factor as well.

\end{document}